\documentstyle[preprint,aps,prl]{revtex}
\begin{document}
\bibliographystyle{prsty}
\draft
%\preprint{\today}
\title{Ising Expansion for the Hubbard Model}
\author{Zhu-Pei Shi and Rajiv R. P. Singh}
\address{Department of Physics, University of California, Davis, California
9561
6}
\maketitle
\begin{abstract}

We develop series expansions for the ground state properties of the
Hubbard model, by introducing an Ising anisotropy into the Hamiltonian. For the
two-dimensional (2D) square lattice half-filled Hubbard model, the ground state
energy,
local moment, sublattice magnetization, uniform magnetic susceptibility
and spin stiffness
are calculated as a function of $U/t$, where $U$ is the Coulomb constant and
$t$ is the hopping parameter.
Magnetic susceptibility data indicate a crossover around $U\approx 4$
between spin density wave antiferromagnetism and Heisenberg antiferromagnetism.
Comparisons with Monte Carlo simulations, RPA result and mean field solutions
are also made.

\end{abstract}

\narrowtext
%%%%%%%%%%%%%%%%%%%%%%%%%%%%%%%%%%%%%%%%%%%%%%%%%%%%%%%%%%%%%%%%%%%%%%%%%%%%%
%%%%%%%%%%%%%%%%%%%%%%%%%%%%%%%%%%%%%%%%%%%%%%%%%%
\section{Introdution}
\hspace*{0.6cm}
Low temperature properties of strongly correlated electron systems have been
studied extensively over the past few decades.
The Hubbard model is one of the simplest models of interacting electrons,
which is believed to capture the main features of electron
correlations in condensed matter systems.
The model  is best understood at half filling, where it exhibits the
Mott-Hubbard
phenomena, namely the existence of insulating behavior accompanied by magnetic
order
in a half-filled band of electrons.
Recent discovery of high-$T_{c}$ superconductivity in doped antiferromagnets
(copper oxides) has led Anderson~\cite{and:88} to suggest that
superconductivity in these materials is closely related to the physics of
strong correlations contained in the Hubbard model.
The spin-half Heisenberg model (a large U limit of the Hubbard model) has
formed
the basis for a quantitative understanding of the magnetic excitations
in the insulating stoichiometric phase of the high-$T_{c}$ materials.
Studies of the Hubbard and $t-J$ models have also provided qualitative insights
into the magnetic and
transport properties of doped high-$T_{c}$ materials~\cite{dag:94}.

The Hubbard model is defined by the lattice Hamiltonian
\begin{equation}
H=-t\;\sum _{<i,j>,\sigma }\; (c^{+}_{i\sigma }\,c_{j\sigma }
                               +c^{+}_{j\sigma }\,c_{i\sigma })
  +U\sum _{i}\;(n_{i\uparrow }-\frac{1}{2})\,(n_{i\downarrow }-\frac{1}{2})
\;-\; \mu \sum _{i} \;(n_{i\uparrow }+n_{i\downarrow })\;\;\; ,
\end{equation}
where, as usual, $c^{+}_{i\sigma }$ and $c_{i\sigma }$ are the creation and
annihilation operators for electrons with a $z-$component of spin $\sigma $ at
lattice site $i$,
and $n_{i,\sigma }=c^{+}_{i\sigma }\,c_{i\sigma }$.
$U$ is the on site repulsive interaction, $\mu $ the chemical potential,
and $t$ the nearest-neighbor hopping amplitude. The sum $<i,j>$ is over all
pairs of
nearest neighbor lattice sites.

In spite of its apparent simplicity, the Hubbard model is exactly soluble
only in one-dimension~\cite{lieb:65}.
In two-dimensions, relevant to the high-$T_c$ materials, this model has been
studied by a variety of approximate methods.
These include finite size diagonalization~\cite{dag:94}, mean-field
theory~\cite{penn:66}, Green's-function decoupling schemes~\cite{hub:64},
variational approaches~\cite{gutz:65},
the random phase approximation (RPA)~\cite{rpa:89}, high temperature series
expansion~\cite{ht:rajiv} and quantum Monte Carlo
simulations~\cite{hir:89,whi:89,mor:90}.
Despite these efforts some of the very basic features of this model, such as
its
phase diagram in the space of temperature, density and $U/t$ are not fully
known.
The numerical studies do provide a consistent picture for the magnetically
ordered zero temperature phase at half-filling.
 However, away from half-filling the possibility
of long range incommensurate magnetic order and superconductivity remain open.

In this paper we develop a series expansion method for studying the Hubbard
model,
and calculate the
ground state energy $E_{g}$, local moment $L$, sublattice magnetization
$M^{+}$,
the uniform magnetic suscebtibility $\chi _{\perp }$ and
spin stiffness $\rho _{s}$,
for the half-filled case, as a function of $U/t$.
This study is complimentary to the existing numerical methods and provides
additional
insight into this model.
Our results are in broad agreement with the Monte Carlo simulations. There are
however,
some quantitative differences. For example the manner in which the local moment
$L$
goes to $1/2$ as $U\to 0$, is different in our calculations than in the
Monte-Carlo
simulations. Our results are much closer to the spin-density-wave mean field
results
in this limit.
Furthermore, the behavior of magentic susceptibility as a function of $U/t$
changes character at $U/t\approx 4$ indicating a crossover between
spin density wave (SDW) antiferromagnet and a Heisenberg antiferromagnet at
this value

The organization of the paper is as follows:
In Sec. II we discuss our series expansion method for the two-dimensional
Hubbard model.
In Sec. III we present results for the ground state properties
and in Sec. IV  for the uniform magnetic susceptibility of the model.
In Sec. V we derive and compute the spin stiffness of the Hubbard model
and estimate the spin wave velocity from the relation
$v^{2}_{s}=\rho _{s}/\chi _{\perp }$.
Finally, in Sec. VI, we summarize our results.

\section{Ising Expansion for the Hubbard Model}

In order to develop series expansion for the Hubbard model, we need to
introduce an Ising
anisotropy into the Hubbard Hamiltonian:
\begin{equation}
H_{\lambda }=-\lambda \,t\;\sum _{<i,j>,\sigma }\; (c^{+}_{i\sigma
}\,c_{j\sigma }
                               +c^{+}_{j\sigma }\,c_{i\sigma })
  +U\sum _{i}\;(n_{i\uparrow }-\frac{1}{2})\,(n_{i\downarrow }-\frac{1}{2})
\;+f(\lambda ) \;\sum _{<i,j>}\;\sigma ^{z}_{i}\, \sigma ^{z}_{j}\;\;\;\; ,
\end{equation}
where $\sigma ^{z}_{i}=(n_{i\uparrow }-n_{i\downarrow })$ is the $z$ component
of the spin at site $i$. The particle-hole symmetry ensures half-filling.
The function $f(\lambda )$ needs to be chosen such that $f(0\leq \lambda <1)
>0$, and $f(1)=0$.
For $\lambda =0$ the atomic limit of the Hubbard model is highly degenerate,
however the Ising term selects from these the N\'eel states as the two
degenerate ground states. Futhermore, this term also introduces a gap in the
spectrum at $\lambda =0$.
{\em For $\lambda =1$ the Ising anisotropy goes to zero and the conventional
Hubbard Hamiltonian is recovered\/}. Ground state properties of the model for
$\lambda \neq 1$ can be obtained by an expansion in powers of $\lambda $.
If the gap does not close before $\lambda =1$ as expected for this model, we
can obtain properties of the Hubbard model by extrapolating the expansions to
$\lambda =1$.

We choose the function $f(\lambda )$ to have the form
\begin{equation}
f(\lambda )=J(1-\lambda )\;\;\; ,
\end{equation}
where $J$ is a constant which can be tuned to improve the convergence of the
extrapolations. In the strong coupling limit, the half-filled Hubbard model is
equivalent to an antiferromagnetic Heisenberg model with Hamiltonian:
\begin{equation}
H_{Heisenberg}=J_{H}\; \sum _{<i,j>}\; (\sigma _{i}\cdot \sigma _{j}\;-\;
1)\;\;\; ,
\end{equation}
where $<i,j>$ denotes nearest neighbors and $J_{H}=t^{2}/U$.
If we choose $J$ in Eq.(3) to be much large than $J_{H}$ we find that the
expansion coefficients become very large and oscillate rapidly, making
extrapolations very difficult. On the other hand, if $J$ is much smaller than
$J_{H}$, the expansion coefficients become very
small and a large number of terms will be needed to extrapolate to $\lambda
=1$. The optimum value of $J$ is around $J_{H}$.

In order to employ series expansion techniques, we write the Hamiltonian in
Eq.(2) as
\begin{equation}
H_{\lambda }=H_{0}\;+\;H^{'}\;\;\; ,
\end{equation}
where unperturbed part of Hamiltonian $H_{0}$ is defined as
\begin{equation}
H_{0}=U\sum _{i}\;(n_{i\uparrow }-\frac{1}{2})\,(n_{i\downarrow }-\frac{1}{2})
+J\;\sum _{<i,j>}\;\sigma ^{z}_{i}\, \sigma ^{z}_{j}\;\;\; ,
\end{equation}
and the perturbation $H^{'}$ is
\begin{equation}
H^{'}=-\lambda \,t\;\sum _{<i,j>,\sigma }\; (c^{+}_{i\sigma }\,c_{j\sigma }
                               +c^{+}_{j\sigma }\,c_{i\sigma })\;
-\; \lambda \,J\;\sum _{<i,j>}\;\sigma ^{z}_{i}\, \sigma ^{z}_{j}\;\;\; .
\end{equation}
Throughout this paper, we take $t=1$ as the energy unit. $U$ and $J$ are two
interaction parameters.
The parameter $\lambda $ simply allows us to extrapolate from the Ising
model to the half-filled Hubbard model.

\section{Ground State Properties}

The ground state energy $E_{g}$ is defined as
\begin{eqnarray}
E_{g}&=&<\psi _{g}(\lambda )\mid H(\lambda ) \mid \psi _{g}(\lambda )>/<\psi
_{g}(\lambda )\mid \psi _{g}(\lambda )> \nonumber \\
 &=&\sum _{n}\; e_{n}(U,J)\;\lambda ^{n} \;\;\; ,
\end{eqnarray}
where $\mid \psi _{g}(\lambda )>$ is the ground state.

The sublattice magnetization $M^{+}$ is defined as
\begin{eqnarray}
2M^{+}&=&\frac{1}{N}\; <\sum _{i} \epsilon (i)\;(n_{i\uparrow }-n_{i\downarrow
})>
\nonumber \\
&=&\sum _{n}\; m_{n}(U,J)\;\lambda ^{n}
\;\;\; ,
\end{eqnarray}
where $N$ is the number of lattice sites, $\epsilon (i)$ set to be $+1$ on one
sublattice and $-1$ on the other,
and the angular brackets refer to ground state expectation values.

The squared local moment $L$ is defined via the relation
\begin{eqnarray}
L&=&\frac{1}{N}\; <\sum _{i} \;(n_{i\uparrow }-n_{i\downarrow })^{2}
\nonumber \\
&=&\sum _{n}\; l_{n}(U,J)\;\lambda ^{n}
\;\;\; .
\end{eqnarray}
The expansion coefficients for $E_{g}$, $L$, and $2M^{+}$ to $11th$ order in
 $\lambda $ are listed in tables I-III.

We find that properties which are not sensitive to long range order, such as
ground state energy $E_{g}$ and local moment $L$ can be easily obtained to high
accuracy with the simplest extrapolations.
Pad\'e approximants~\cite{hun:79} to the function under consideration may be
formed directly.
In order to estimate
the values of $E_{g}$ and $L$ at particular $U$, we compute all
Pad\'e approximants from series up to $9th$, $10th$ and $11th$
orders for three different
Ising parameter $J/(t^{2}/U)=0.75,1.0,1.25$,
and remove ill-behaved P\'ade approximants.
The estimates are set to be the median of
the P\'ade approximants at $\lambda =1$, and the errorbars are estimated as
half the spread of these P\'ade approximants.

The values of the ground state energies at $\lambda =1$
are $E_{g}=-1.51(2), -1.86(1), -2.934(4)$ and $-5.2293(8)$ for
$U=1, 4, 10$ and $20$, respectively.
They are plotted in Fig.~1 as solid circles with errorbars. The solid line
is a guide to the eye. Our calculated ground state energies $E_{g}$ agree
with the Monte Carlo simulations.
For example, at $U=4$, the extrapolated ground state energy from
finite temperature Monte Carlo simulations~\cite{whi:89} is $-1.864(1)$.
Projected Monte Carlo simulations~\cite{ima:92}
(an algorithm which allows the calculation of ground state properties
in the canonical ensemble, {\em i.e.}, with a fixed number of electrons)
 gives $-1.864(2)$.
At $U=1$, our result ($E_{g}+U/4=-1.26$) is close to the slave-boson mean field
result ($\approx -1.3$)~\cite{lil:90}

The squared local moment obtained from the P\'ade approximants at $\lambda =1$
are
$L=0.54(1), 0.75(1), 0.924(2)$ and $0.9788(4)$ for $U=1,4,10$ and $2
0$, respectively. They are plotted in Fig.~2 as solid circles with errorbars.
Once again, the solid line is a  guide to the eye. The empty squares and dashed
lines are the
Monte Carlo simulations obtained by White {\em et al.\/}~\cite{whi:89}
at low temperatures [Fig.~6(a) in Ref.~9]. We see that both calculations
are in good agreement with each other at large $U$ but they show
different behaviors at small $U$.
 Monte Carlo simulations show that local moment $L$ increases
rapidly as $U$ increases
({\em concave function\/}) while ours indicate  a relatively slow
growth ({\em convex function \/}) at small $U$ region, as shown in Fig.~3.
We would naively expect the series extrapolation to be better at large $U$, and
not so well at small $U$. However, as we now show, our results are rather
close to the weak coupling results at small $U$, which suggest that our
convergence is reasonable for all $U$.

Within the Hartree-Fock approximation, we can derive a simple relationship
between
local moment and sublattice magnetization.
The local moment is
\begin{equation}
L=<n_{\uparrow }>\; + \; <n_{\downarrow }>\;-\;2\, <n_{\uparrow
}><n_{\downarrow }>\;\;\; ,
\end{equation}
where
\begin{eqnarray}
<n_{\uparrow }>&=& 0.5\;+\; M^{+} \nonumber \\
<n_{\downarrow }>&=& 0.5\;-\; M^{+}\;\;\; .
\end{eqnarray}
Substituting Eq.(12) into Eq.(11) we have
\begin{equation}
L=2\,(M^{+})^{2}+0.5\;\;\; ,
\end{equation}
The mean field solution for
the sublattice magnetization is~\cite{rpa:89,hir:85}
\begin{equation}
M^{+}=\Delta /U \;\;\; ,
\end{equation}
where $\Delta $ can be obtained by solving the gap equation
\begin{equation}
\frac{U}{2N}\; \sum _{{\bf k}} \; \frac{1}{(\epsilon ^{2}_{{\bf k}}+\Delta
^{2})^{1/2}}\;=\; 1\;\;\; .
\end{equation}
$N$ is the total number of sites and
$\epsilon ({\bf k})=-2t\,(cos\,k_{x}+cos\,k_{y})$. The local moment given by
numerical solution of the gap equation is plotted in Fig.~2, a dashed line.
One can see that the function $L(U)$ has  a convex shape at small $U$ which
agrees with our conclusion.
At $U=1$, our result ($L=0.54$) agrees with the slave-boson mean field result
($L\approx 0.53$)~\cite{lil:90}

The extrapolation of sublattice magnetization $M^{+}$ appears much harder.
In the 2D spin-$\frac{1}{2}$ square lattice Heisenberg antiferromagnet,
the sublattice magnetization ($2M^{+}$) is estimated by expansions
around the Ising limit. In order to extrapolate the $2M^{+}$ series reliably,
one has to remove the singularity of the form
$[1-(J_{\perp }/J_{\parallel })^{2}]^{\frac{1}{2}}$ caused by
Goldstone modes~\cite{sin:89}, where $J_{\perp }$ is the exchange
perpendicular to the direction
of ordering and $J_{\parallel }$ the exchange along it.
One way of doing this is
to go a new variable $\delta $ given by~\cite{hus:88}
\begin{equation}
1-\delta =[1-(J_{\perp }/J_{\parallel })^{2}]^{\frac{1}{2}}\;\;\; .
\end{equation}
We expect to have the same situation for the sublattice magnetization $2M^{+}$
in the 2D Hubbard model because it is equivalent to the Heisenberg model at
large $U$. Here we set a new varible
$\delta $ to be
\begin{equation}
1-\delta =(1-\lambda )^{\frac{1}{2}}\; \; \; ,
\end{equation}
where $\delta =1$ at $\lambda =1$.
Then the series for the sublattice magnetization $2M^{+}$ becomes
\begin{equation}
2M^{+}=\sum ^{11}_{n=0}\; b_{n}(U,J)\, \delta ^{n} \;\;\; ,
\end{equation}
where $b_{n}(U,J)$ can be  calculated from previous series $m_{n}(U,J)$.
Again, we compute all Pad\'e approximates from 7th order to 11th order series.
The median of Pad\'e approximants at $\delta =1$ give $2M^{+}=0.12(5), 0.37(5),
0.58(3)$, and $0.61(3)$ for $U=1, 4, 10$ and $20$, respectively.
They are plotted in Fig.~3, solid circles with errorbars.
The solid line is a guide to the eye.

At large $U$, the
effective Hamiltonian for the half-filled Hubbard model is known to be the
spin-$\frac{1}{2}$ Heisenberg model.
The quantum zero-point spin wave fluctuations reduce the sublattice
magnetization of $2M^{+}$ from the classical value of unity to about
$2M^{+}=0.6$ as first obtained by Anderson~\cite{and:52}. This
is also supported by Monte Carlo simulations~\cite{you:88} and
series expansion
studies~\cite{sin:89}. At $U=20$, we estimate $2M^{+}=0.61$ which is
close to the value $0.6$ of the
Heisenberg model.
Also, at $U=4$, we can compare our results [$2M^{+}=0.37(5)$] with the Monto
Carlo simulations [$2M^{+}=0.39(5)$] of Hirsch and Tang~\cite{hir:89} and find
fairly good agreement.

Also, we compare our results with the mean field solutions and the RPA
calculations.
The sublattice magnetization given by
numerical solution of the gap equation is plotted in Fig.~3,
as a long dashed line.
At large $U$, the mean field solution gives $2M^{+}=1$, which is far
from the true value ($2M^{+}=0.6$) of the Heisenberg model
because of the neglect of the quantum fluctuations.
In order to take into account the fluctuation effects on the sublattice
magnetization, one simply has to calculate the self-energy correction to the
one
particle Green's function as done by Schrieffer, Wen and Zhang~\cite{rpa:89} in
random phase approximation. Their RPA result is plotted in Fig.~3 by
 a dotted line.
Clearly, the RPA result brings the mean field result much closer to
our numerical estimates.

We note again that the purpose of introducing the Ising anisotropy $J(1-\lambda
)\sigma ^{z}_{i}\;
\sigma ^{z}_{j}$ into the conventional Hubbard Hamiltonian is to be able to
carry out the series
expansions. The ground state properties of the Hubbard model should not depend
on the parameter $J$ of the Ising anisotropy.
One can also choose another type of function $f(\lambda )$ in Eq.(2):
\begin{equation}
f(\lambda )=J\;(1-\lambda ^{2})\;\;\; .
\end{equation}
This form of $f(\lambda )$ leads to an expansion in powers of $\lambda ^{2}$.
This allows one to treat the $J$ term generated by the perturbation in the
Hubbard model and the Ising term introduced by hand on {\em an equal
footing\/}.
However, it has the disadvantage that the number of series coefficients
available
for extrapolation is reduced by half. The resulting extrapolations at $\lambda
=1$
are consistent with those obtained earlier.

\section{Magnetic Susceptibility}

We now discuss the behavior of the uniform magnetic susceptibility of the 2D
Hubbard model from the series expansions. We turn on the magnetic field and
add a Zeeman interaction into the 2D Hubbard Hamiltonian:
\begin{eqnarray}
H_{\lambda h}&=&-\lambda \,t\;\sum _{<i,j>,\sigma }\; (c^{+}_{i\sigma
}\,c_{j\sigma }
                               +c^{+}_{j\sigma }\,c_{i\sigma })
  +U\sum _{i}\;(n_{i\uparrow }-\frac{1}{2})\,(n_{i\downarrow
}-\frac{1}{2})\nonumber \\
& &+J(1-\lambda ) \;\sum _{<i,j>}\;\sigma ^{z}_{i}\, \sigma ^{z}_{j}
\; -\; h\, \sum _{i} \; \sigma ^{x}_{i} \;\;\;\; ,
\end{eqnarray}
where $h$ is magnetic field and $\frac{1}{2}\sigma ^{x}_{i}$ is the
$x-$component of the  spin at site $i$. The uniform magnetic
susceptibility $\chi _{\perp }$ is defined as
\begin{eqnarray}
4\chi _{\perp }&=&-\partial ^{2}E/\partial h^{2}\; \mid _{h=0} \nonumber \\
&=& \sum _{n}\; x_{n}(U,J)\;\lambda ^{n}\;\;\; .
\end{eqnarray}
The expansion coefficients of $4\chi _{\perp }$ to 9th order in $\lambda $ are
given in Table IV.

At half-filling, 2D Hubbard model has antiferromagnetic long range order.
One can see that $\chi _{\perp }$ series is dominated by a simple pole at
$\lambda =-1$.
We need to remove this singularity before further analysis. This
is done by going to a new variable $\beta $ given by
\begin{equation}
\beta =2\lambda /(1+\lambda ) \;\;\; .
\end{equation}
The resulting series has the form
\begin{equation}
4\chi _{\perp }=\sum ^{9}_{n=0} \;y_{n}(U,J)\;\beta ^{n}\;\;\; ,
\end{equation}
where $y_{n}(U,J)$ can be obtained from the previous series $x_{n}(U,J)$. But
this new series still contains singularities at the Hubbard point ($\lambda =1$
or $\beta =1$). Going to a new variable $\delta $ defined by
\begin{equation}
1-\delta =(1-\beta )^{1/2} \;\;\; ,
\end{equation}
and constructing Pad\'e approximants, we find that only two Pad\'e approximants
($K=4, M=5$ and $K=5, M=4$)
from 9th order series show good behavior and both give the same results.
For this reason it is difficult to estimate errors in these results.
The uniform magnetic susceptibility obtained from these Pad\'e approximants
are $4\chi _{\perp }=0.94, 0.58, 0.75, 0.92$ and $1.40$
for $U=1, 4, 6, 10$ and $20$, respectively.
They are plotted in Fig.~4 as  solid circles. The solid line is a  guide to
the eye.

At $U=20$, $\chi _{\perp }/U=\chi _{\perp }\,J=0.0175$ is close to the
large $U$ limit value $\chi _{\perp }\,J_{H}=0.0163$ of
the Heisenberg
model~\cite{sin:89} shown by a dashed line in Fig.~4(a).
At $U=4$, our result ($4\chi _{\perp }=0.58$) agrees with the estimated value
($\approx 0.53$) of Monte Carlo simulations
at low temperature by White {\em et al.\/}~\cite{whi:89}, but does not agree
with the estimated value ($\approx 1.0$) of projected Monte Carlo simulations
by Furukawa and Imada~\cite{ima:92}.
At $U=10$, our result ($4\chi _{\perp }=0.92$) agrees with the estimated value
($\approx 0.98$) of Monte Carlo simulations by Moreo~\cite{mor:93}.

One can see that our magnetic susceptibility data indicate a dip at
$U=U_{c}\approx 4$, as shown in the Fig.~4(b).
It suggests that there is a crossover around $U_{m}$ in the behavior of the 2D
Hubbard model at half filling.
For $U>U_{c}$, we can see that magnetic susceptibility ($4\chi _{\perp }$) is
almost proportional to $U$ or inversely proportional to the spin
superexchange $J$.
This can be taken as evidence
that the magnetic state for $U>U_{c}$ for 2D Hubbard model is that of
the Heisenberg model. But, for
$U<U_{c}$, magnetic susceptibility ($4\chi _{\perp }$) has a different
behavior, and
decreases as $U$ increases. This can be interpreted as a weak coupling
behavior of the SDW ground state.
Thus we obtain a crossover between these types of behavior around $U\approx 4$.

Also, we compare our result with the mean field solutions of uniform magnetic
susceptibility~\cite{den:93}:
\begin{equation}
\chi _{\perp }=\frac{1}{2U}\;[(1-\Delta
^{2}\,U\,I_{1}-U\,I_{2}^{2}/I_{1})^{-1}\
;-\;1] \;\;\; ,
\end{equation}
where
\begin{eqnarray}
I_{1}&=&\frac{1}{2N}\; \sum _{{\bf k}}\; \frac{1}{E^{3}({\bf k})} \nonumber \\
I_{2}&=&-\frac{1}{2N}\; \sum _{{\bf k}}\; \frac{\epsilon ({\bf k})}{E^{3}({\bf
k
})} \nonumber \;\;\;.
\end{eqnarray}
Here $E({\bf k})=[\epsilon ^{2}({\bf k})+\Delta ^{2}] ^{1/2}$.
It is plotted in Fig.~4(b), a solid line. We see that the mean field result
is
qualitatively correct, but overestimates $\chi _{\perp }$,
especially at large $U$.

\section{Spin Stiffness of the Hubbard Model}

The spin stiffness constant $\rho _{s}$ is a measure of the response of the
spin
system in an ordered phase to a twist of the order parameter.
To our knowledge, there are no quantum
Monte Carlo estimates for the spin stiffness in the Hubbard model.
Derivation and calculation of this quantity for the Hubbard model was discussed
by us recently in a short communication~\cite{shi:prl}. For completeness,
we discuss it again here.

If we rotate the ordering direction by a small angle $\theta $ along a
given direction such as $y$ axis, then the spin stiffness constant
$\rho _{s}$ can be defined through the increase of the ground state energy:
\begin{equation}
E_{g}(\theta )=E_{g}(\theta =0)\;+\; \frac{1}{2}\, \rho _{s}\,\theta ^{2}\;
+\; O(\theta ^{4})\;\;\; ,
\end{equation}
This rotation can be carried out by the following transformation applied to
the Fermion operators:
\begin{equation}
\left(
\begin{array}{c}
c'_{\uparrow }\\
c'_{\downarrow}
\end{array}\right)\; =\;
\left(\begin{array}{cc}
cos\,\phi /2 & sin\,\phi /2 \\
-sin\,\phi /2  & cos\,\phi /2
\end{array}\right)\;\;
\left(
\begin{array}{c}
c_{\uparrow }\\
c_{\downarrow}
\end{array}\right)
\;\;\; ,
\end{equation}
and
\begin{equation}
\left(
\begin{array}{c}
c'^{+}_{\uparrow }\\
c'^{+}_{\downarrow}
\end{array}\right)\; =\;
\left(\begin{array}{cc}
cos\,\phi /2 & sin\,\phi /2 \\
-sin\,\phi /2  & cos\,\phi /2
\end{array}\right)\;\;
\left(
\begin{array}{c}
c^{+}_{\uparrow }\\
c^{+}_{\downarrow}
\end{array}\right)
\;\;\; .
\end{equation}
This transformation generates desired spin rotation:
\begin{eqnarray}
\sigma _{z'}&=&c'^{+}_{\uparrow }c'_{\uparrow }\,-\,
             c'^{+}_{\downarrow}c'_{\downarrow}\nonumber \\
            &=&\sigma _{z}\, cos\,\phi +\sigma _{x}\, sin\, \phi \nonumber \\
\sigma _{x'}&=&c'^{+}_{\uparrow }c'_{\downarrow}\,-\,
               c'^{+}_{\downarrow}c'_{\uparrow }\nonumber \\
             &=&-\sigma _{z}\, sin\, \phi +\sigma _{x}\,cos\,\phi
\end{eqnarray}
After rotation by a {\em relative\/} angle $\theta $
( that is letting $\phi$ change by $\theta/2$ ) between neighboring
sites separated along $y$ axis ($\hat y$ denotes unit distance in $y$
direction)
,
$H$ in Eq.(1) becomes
\begin{equation}
H_{\theta }=H\;+\; H^{dia}\;+\; H^{para} \;+\;O(\theta ^{3})\;\;\; ,
\end{equation}
where
\begin{eqnarray}
H^{dia}&=&\frac{t\theta ^{2}}{8}\;\sum _{i,\sigma }\; (c^{+}_{i\sigma }
                                  \,c_{i+\hat{y}\sigma }
                               +c^{+}_{i+\hat{y}\sigma }\,c_{i\sigma
})\nonumber
\\
H^{para}&=& -\frac{t\theta }{2}\;\sum _{i}\; (c^{+}_{i\uparrow }
\,c_{i+\hat{y}\downarrow }\;-\; c^{+}_{i\downarrow }c_{i+\hat{y}\uparrow }
\nonumber \\
& &\phantom{ \frac{t\theta }{2}\;\sum _{<i,\hat{y}>}\; (}
   \;+\; c^{+}_{i+\hat{y}\downarrow } c_{i\uparrow }\;-\;
c^{+}_{i+\hat{y}\uparrow }c_{i\downarrow })\;\;\; .
\end{eqnarray}
The ``diamagnetic" term $H^{dia}$ is already of order $\theta ^{2}$ so
for the calculation of the energy to order $\theta^2$, it
can be replaced by its expectation value in the ground state of the $\theta =0$
Hamiltonian. We have
\begin{eqnarray}
\rho ^{dia}_{s}&=&-\frac{1}{8}\; (Kinetic~ Energy) \nonumber \\
  &=&-\frac{1}{8}\;[E_{g}(\theta =0)-\frac{U}{2}\;(n-L)] \;\;\; ,
\end{eqnarray}
where $n$ is a band filling, and $L$ a local moment defined in Eq.(10).
We have already obtained $E_{g}$ and $L$ in Sec. III,
and $n=1$ for half-filling.

The contribution of the ``paramagnetic" term $H^{para}$ to the ground state
energy in order $\theta ^{2}$ can be obtained by treating it in
second order perturbation theory. We have
\begin{eqnarray}
\rho ^{para}_{s}&=&2\frac{\partial ^{2}E}{\partial \theta ^{2}}\mid _{\theta
=0} \
;\;\; \nonumber \\
&=& \sum _{n}\;p_{n}\;\lambda ^{n}
\;\;\; ,
\end{eqnarray}
where $E$ represents energy of the Hamiltonian $H_{\lambda }+H^{para}$.
The series coefficients $p_{n}$ of $\rho ^{para}_{s}$ to
$9th$ order in $\lambda $
are given in the Table V.

We use the same Pad\'e analysis for $\rho ^{para}_{s}$ as for
uniform magnetic susceptibility.
We obtain the spin stiffness
$\rho _{s}=\rho ^{dia}_{s}+\rho ^{para}_{s}=0.186(15), 0.15(1), 0.077(7)$
and $0.039(5)$ for
$U=1, 4, 10$ and $20$, respectively.
They are plotted in Fig.~5 as filled circles with a dashed line
as  guide to the eye.
The errorbars represent the spread in the Pad\'e estimates.
At $U=20$, our result ($\rho _{s}U=0.78$) agrees well with the
value ($0.73$) of the Heisenberg model~\cite{sin:89} shown by
a short solid line.

We also compare our results with the
Hartree-Fock approximation for the spin stiffness~\cite{den:93}:
\begin{equation}
\rho _{s}=-\frac{t}{2N}\;\sum _{{\bf k}}\; \frac{\epsilon ({\bf k})\;
           cos\,k_{x}}{E({\bf k})} \;\;\; .
\end{equation}
The spin stiffness given by mean field solution of the gap equation is
plotted in Fig.~5(b) as a solid line. One can see that the mean field
approximation overestimates the stiffness at large $U$.
At $U=20$, the mean field result is $\rho _{s}U=0.98$ compared to
$0.73$ for the Heisenberg model.

We note that the spin stiffness (filled squares in Fig.~1) from
the variational Monte Carlo method
with a Gutzwiller-type wave function are even larger than values of
the mean field solution~\cite{den:93}.
We believe that the large discrepancy is due to the missing spin flip
processes in the Gutzwiller variational wave function used in ~\cite{den:93}.
Their calculations only get contributions to the spin-stiffness
beyond the Hartree-Fock result from the ``diamagnetic'' term,
whereas the ``paramagnetic" part of the spin stiffness which contains
spin-flip processes does not get corrected.
The expectation value of the operator $H^{para}$ between the Hartree-Fock and
Gutzwiller wave functions is zero~\cite{den:93b}.
This suggests that one might improve the variational Monte Carlo method
by adding a spin flip operator $O_{sp}$ into the Gutzwiller-type wave function:
\[
\mid \Psi _{GW}>\;=\;\prod _{i}\; [1-(1-g)n_{i\uparrow }\,n_{i\downarrow }
+p\times O_{sp}]\;\mid \Psi _{HF}> \nonumber
\]

where $p$ is a parameter and
\[
O_{sp}=(c^{+}_{i\uparrow }
\,c_{i+\hat{y}\downarrow }\;-\;
c^{+}_{i\downarrow }c_{i+\hat{y}\uparrow }
   \;+\; c^{+}_{i+\hat{y}\downarrow } c_{i\uparrow }\;-\;
c^{+}_{i+\hat{y}\uparrow }c_{i\downarrow })\nonumber \;\;\; .
\]

We now turn to the calculation of spin-wave velocity. This can be calculated
analytically by adding RPA fluctuations to the SDW state~\cite{rpa:89}.
Designed, at first sight, for intermediate and weak coupling ($U/4t \leq 1$)
 this method also interpolates smoothly to
the Heisenberg limit at large $U/t$.

Alternatively, there is a relationship between the spin stiffness $\rho _{s}$
of
the order parameter with respect to a spiral twist, the spin wave velocity
$v_{s
}$ and the uniform magnetic susceptibility $\chi _{\perp }$ transverse to the
stagged spin density:
\begin{equation}
\rho _{s}=v_{s}^{2}\; \chi _{\perp }\;\;\; .
\end{equation}
It was originally derived by the hydrodynamic theory of Halperin and Hohenberg
{}~\cite{hal:69}.
The spin wave velocity obtained from this relationship
($\rho _{s}=v^{2}_{s}\chi _{\perp }$) are
$v_{s}=0.89,~1.02,~0.58$ and $0.34$ for
$U=1,~4,~10$ and $20$, respectively.
They are plotted in Fig.~6 as filled circles with a solid line
as a guide to the eye. At $U=20$, our $v_{s}=0.34$ agrees well with
the limiting value ($v_{s}=1.18\sqrt{2}\,J'=0.33$) of
the Heisenberg model~\cite{sin:89}, a dashed line in Fig.~6,
where $J'=4t^{2}/U$.

We compare this result with the RPA solution of
Schrieffer, Wen and Zhang~\cite{rpa:89}:
\begin{equation}
v_{s}=[4t^{2}(1-\Delta ^{2}U\,I_{1})\;\frac{I_{3}}{I_{1}}]^{1/2}\;\;\; ,
\end{equation}
where
\begin{equation}
I_{3}=\frac{1}{2N}\; \sum _{{\bf k}}\; \frac{sin^{2}\,(k_{x})}{E^{3}({\bf k})}
\;\;\; . \nonumber
\end{equation}
It is plotted in Fig.~6 as a dotted line. We see that the RPA result for
the spin wave velocity are lower than our series expansion results.

\section{Summary}

To summarize, in this paper we developed series expansions for the half-filled
Hubbard model,
by introducing an Ising anisotropy into the conventional Hubbard Hamiltonian.
Series were developed for
the ground state energy, the local moment, the sublattice magnetization,
the uniform magnetic susceptibility and the spin stiffness,
and were extrapolted to the Hubbard model
using standard series extrapolation methods.
In general, our results are in good agreement with the quantum Monte Carlo
simulations.
There are some quantitative differences, which are also discussed.
Our magentic susceptibility data indicate a crossover between
SDW antiferromagnetic state at small $U$ and
a Heisenberg antiferromagnet at large $U$ at $U\approx 4$.
We note that our results for $\rho_s$
as well as $M^\dagger$ are somewhat higher than the known results for
the Heisenberg models. We believe this reflects the fact that the
series have not converged as well as for the Heisenberg model and hence
the reduction in these quantities due to the zero-point
spin-wave fluctuations
is not fully accounted for. Still, the convergence ($\sim 5\% $)
is quite reasonable.

This research was supported in part by NSF (DMR-9318537) and
University Research Funds of the University of California at Davis.
Zhu-Pei Shi would like to thank Prof. Richard T. Scalettar and
Prof. Barry M. Klein for discussions.

\pagebreak

\pagebreak
{
\widetext
\squeezetable
\begin{table}
\caption{Expansion coefficients for the ground state energy.
}\label{t-tb1}
\begin{tabular}{ccl}
 $U$    &$ J/(t^{2}/U)$    &\multicolumn{1}{c}{$e_{n}(U,J)$, $n= 0 - 11$}\\
\tableline
1  & 0.75&  -3.25~ 3.0~      -0.64~     -0.5376~     -0.36882596~
-0.17599739~     -0.0037101999~  0.11731865~    \\
  &   &  0.17602728~      0.17413953 ~     0.11347540~     -0.0088559404~\\
 1 & 1.0 & -4.25~ 4.0~ -0.5~ -0.4375~ -0.34273695~ -0.23157505~ -0.12102132~
-0.025358881\\
   &     & 0.046126082~ 0.089325558~ 0.10390817~ 0.091603525\\
1  & 1.25& -5.25~ 5.0~     -0.41025641~     -0.36817883~     -0.30808094~
-0.23717612~     -0.16320812~\\
 &  & -0.093237945~ -0.032792472~  0.014501216~  0.046805520~  0.063642701\\
\tableline
4  & 0.75&  -1.75~ 0.75~     -0.75294118~     -0.18602076~     0.032575733~
-0.11521776~      0.051494035~  \\
 &  &   0.41270231~   0.36649252~     -0.23825557~     -0.77368699~
-0.75051947~\\
4  & 1.0& -2.0~ 1.0~ -0.69565217~ -0.21172023~ -0.030847663~
-0.060114049~ -0.011016385~ \\
 &  &  0.16603971~ 0.22775615~ 0.035681298~ -0.22467766~ -0.32023724\\
4  & 1.25&   -2.25~ 1.25~     -0.64646464~     -0.22854811~     0.042172811~
-0.035358477~ -0.023047559~  \\
 &  & 0.066640645~ 0.13361749 ~     0.080005270~ -0.048764273~ -0.13678257\\
\tableline
10  & 0.75&  -2.8~ 0.3~     -0.38004751~     -0.018957239~ 0.066441099~
-0.087012637~  0.038082237~ \\
 &  &   0.16089588~      0.11974079~     -0.075315493~ -0.24874869~
-0.20497804~\\
10  & 1.0& -2.9~ 0.4~ -0.37383178~ -0.024456284~ -0.045510051~
 -0.064243043~ 0.0053501507~ \\
 &  & 0.075910665~ 0.074634170~ 0.0084898042~ -0.068321715~ -0.096846836~\\
10  & 1.25& -3.0~ 0.5~     -0.36781609~     -0.029594398~ 0.033584424~
-0.050422135~ -0.0070731834~ \\
 &  &  0.039086550~  0.046846607~  0.021090155~ -0.015763853~ -0.040974167\\
\tableline
20  & 0.75& -5.15~ 0.15~ -0.19740901~     -0.0025574270~ 0.041371562~
-0.044384097~  0.016313799~  \\
 &  & 0.075735746~  0.056531614~ -0.041034781~ -0.11826259~     -0.071871255~\\
20  & 1.0& -5.2~ 0.2~ -0.19656020~ -0.0033806422~ -0.030305728~
 -0.033253249~ 0.0010839733~\\
 & & 0.034576709~ 0.03478823~ 0.019973437~ -0.033086550~ -0.037873386~ \\
20  & 1.25& -5.25~ 0.25~ -0.19571865~ -0.0041896960~ -0.023690877~
 -0.02656723~ -0.0044976664~\\
 &  &  0.016991751~ 0.021350382~ 0.0088434956~ -0.0081635325~ -0.016372525~\\
\end{tabular}
\end{table}
}

{
\widetext
\squeezetable
\begin{table}
\caption{Expansion coefficients for the squared local moment.
}\label{t-tb2}
\begin{tabular}{ccl}
 $U$    &$ J/(t^{2}/U)$    &\multicolumn{1}{c}{$l_{n}(U,J)$, $n= 0 - 11$}\\
\tableline
1  & 0.75& 1.0~ 0.0~ -0.2048~     -0.34406400~     0
.34363876~     -0.18584371~      0.073558744~ \\
& & 0.3
2646019~      0.45962201~      0.39942169~      0.126
77212~     -0.33704683~\\
1  & 1.0& 1.0~ 0.0~ -0.125~ -0.21875~ -0.25342785~ -0.21761601 -0.12058227~\\
& & 0.010355528~ 0.13808712~ 0.22622422~ 0.248
12902~ 0.19219824~\\
1  & 1.25& 1.0~ 0.0~ -0.084155161~ -0.15104772~     0
.18802143~     -0.18850833~     -0.15340922~    \\
& & -0.090723264~ -0.013648193~  0.062253229~  0.12199406~      0.15344711~\\
\tableline
4  & 0.75& 1.0~ 0.0~ -0.28346021~     -0.14006269~     0.020099625~
-0.042208513~  0.13810651~  \\
 & &    0.6
6290690~      0.47876441~     -0.91546939~      -1.99
197718~      -1.29230064~\\
4  & 1.0& 1.0~ 0.0~ -0.24196597~ -0.147283644~ 0.0033966659~ 0.0020465901~
0.041625973~\\
 &  &  0.27003095~ 0.33545751~
 -0.10812405~ -0.68716227~ -0.74045463\\
4  & 1.25& 1.0~ 0.0~ -0.20895827~     -0.14774827~
0.017014379~  0.014267802~  0.020573983~  \\
 & & 0.1
1999820~      0.20351777~      0.071247345~ -0.22059729~     -0.38236644~\\
\tableline
10  & 0.75& 1.0~ 0.0~ -0.072218053~ -0.0072046514~
0.022443265~ -0.034066355~  0.041464469~  \\
 &  & 0.1
1664774~      0.058849733~ -0.12141417~     -0.25211532~     -0.14201779\\
10  & 1.0& 1.0~ 0.0~ -0.069875098~ -0.0091425362~ -0.014415554~
 -0.024755587~ 0.016845440~\\
&  &  0.060456748~
0.045459350~ -0.021365905~ -0.086403185~ -0.092929358~\\
10  & 1.25& 1.0~ 0.0~ -0.067644339~ -0.010885296~
0.0099895925~ -0.019054664~  0.0064590300~ \\
& & 0.035218951~  0.032572341~  0.0029017356~ -0.031199252~ -0.048202015~\\
\tableline
20  & 0.75& 1.0~ 0.0~ -0.019485158~ -0.00050485912~
0.0079729731~ -0.0088647321~  0.0094722236~  \\
 & & 0.027330357~  0.013712645~ -0.031211772~ -0.058546614~ -0.016716436~\\
20  & 1.0& 1.0~ 0.0~ -0.019317955~ -0.00066449970~ -0.0057661935~
-0.0066346164~ 0.0037510047~ \\
& & 0.013829393~
0.010568695~ -0.0058676230~ -0.020309374~ -0.016500150~\\
20  & 1.25& 1.0~ 0.0~ -0.019152896~ -0.00082000166~ -0.0044501566~
-0.0052935510~ 0.013853661~\\
 & &  0.078617712~
0.0074609205~ 0.00033154323~ -0.0074336626~ -0.0092115763~\\
\end{tabular}
\end{table}
}

{
\widetext
\squeezetable
\begin{table}
\caption{Expansion coefficients for the sublattice magnetization.
}\label{t-tb3}
\begin{tabular}{ccl}
 $U$    &$ J/(t^{2}/U)$    &\multicolumn{1}{c}{$m_{n}(U,J)$, $n= 0 - 11$}\\
\tableline
1  & 0.75& 1.0~ 0.0~ -0.2048~     -0.34406400~
0.37832834~     -0.30858617~     -0.16797349~      0.0039694233~\\
&  &  0.17811940~      0.32196832~      0.35932350~      0.14690771~\\
1  & 1.0& 1.0~ 0.0~ -0.125~ -0.21875~ -0.26577581~ -0.26255537~ -0.21639940~
 -0.14085987~ \\
 & & -0.050839444~ 0.039513722~
 0.11520618~ 0.15738379~\\
1  & 1.25& 1.0~ 0.0~ -0.084155161~ -0.15104772~
.19346230~     -0.20866262~     -0.19818728~     -0.16672332~    \\
& & -0.12065102~     -0.066877941~ -0.012424680~  0.035149309~\\
\tableline
4  & 0.75& 1.0~ 0.0~ -0.28346021~     -0.14006269~
0.29094791~     -0.70653657~     -0.11899194~       1.78106763~     \\
& &  2.65518980~      0.17150140~      -4.57860227~      -8.08441904\\
4  & 1.0& 1.0~ 0.0~ -0.24196597~ -0.14728364~ -0.16340074~ -0.36296753~
-0.23038035~ \\
 & & 0.49230029~ 1.09523777~
0.69655017~ -0.61481862~ -2.00113858~\\
4  & 1.25& 1.0~ 0.0~ -0.20895827~     -0.14774827~
0.11819956~     -0.21655277~     -0.20123752~      0.10926874~   \\
&  &   0.47800300~      0.49583051~      0.078098871~ -0.50046276\\
\tableline
10  & 0.75& 1.0~ 0.0~ -0.072218053~ -0.0072046514~
0.38636976~     -0.78484180~     -0.10195808~  \\
&  &     1.63572360~       2.29944530~     -0.28902064~      -4.69346784~
-5.25870098\\
10  & 1.0& 1.0~ 0.0~ -0.069875098~ -0.0091425362~ -0.21519753~
 -0.44065350~ -0.20500969~\\
 & &  0.47694043~ 0.93375402~
 0.48462349~ -0.74441315~ -1.65106094~\\
10  & 1.25& 1.0~ 0.0~ -0.067644339~ -0.010885296~
0.13638406~     -0.28150839~     -0.18799630~    \\
&  &   0.13527913~      0.41216832~      0.36759744~     -0.024358458~
-0.48058650\\
\tableline
20  & 0.75& 1.0~ 0.0~ -0.019485158~ -0.00050485912~
0.39342412~     -0.78866044~     -0.12416041~   \\
& &    1.59373845~       2.17733968~     -0.58972155~      -4.77779001~
-4.11947852~\\
20  & 1.0& 1.0~ 0.0~ -0.019317955~ -0.00066449970~ -0.22095641~
-0.44335646~ -0.21769894~\\
&  & 0.45233967~ 0.87748571~
0.35719439~ -0.84087768~ -1.42111547~\\
20  & 1.25& 1.0~ 0.0~ -0.019152860~ -0.00082000166~ -0.14118233~
 -0.28358286~ -0.19603500~\\
& &  0.11902603~ 0.38038377~
0.30147500~ -0.092194212~ -0.43888627\\
\end{tabular}
\end{table}
}

{
\widetext
\squeezetable
\begin{table}
\caption{Expansion coefficients for the uniform magnetic susceptibility.
}\label{t-tb4}
\begin{tabular}{ccl}
 $U$    &$ J/(t^{2}/U)$    &\multicolumn{1}{c}{$x_{n}(U,J)$, $n= 0 - 9$}\\
\tableline
1  & 1.0 & 0.25~ 0.25~ 0.17261905~ 0.027848648~ -0.1262274~ -0.19962669~
-0.11626869~\\
& &
0.12849035~ 0.43255914~ 0.60788905\\
4  & 1.0 &  1.0~  1.0~  -0.67116442~  -2.82447457~  -1.58009047~  4.43513119~
8.64219154~\\
& &  1.42302418~  -15.39992391~
 -22.54729517~\\
6  & 1.0 &  1.5~       1.5~      -1.48149027~
-4.91068070~      -1.92590553~       8.56642180~       13.73914475~ \\
& &     -1.78852254~      -29.87404955~      -32.12999668~\\
10  & 1.0 & 2.5~ 2.5~ -2.99306647~ -8.80889678~ -2.67462512~ 16.12944762~
23.047960372~\\
& & -8.54214685~ -56.78749173~ -44.88013932\\
20  & 1.0 & 5.0~ 5.0~ -6.48980810~ -18.15418203~ -4.79124584~ 33.98201522~
45.62214165~\\
&  & -23.66505149~ -119.66330645~ -76.12120519\\
\end{tabular}
\end{table}
}

{
\widetext
\squeezetable
\begin{table}
\caption{Expansion coefficients for the ``paramagnetic" part
($\rho ^{para}_{s}$) of the spin stiffness.
}\label{t-tb5}
\begin{tabular}{ccl}
 $U$    &$ J/(t^{2}/U)$    &\multicolumn{1}{c}{$p_{n}(U,J)$, $n= 0 - 9$}\\
\tableline
1  & 1.0 &0.0~ 0.0~ 0.0~ 0.0~ 0.000047115128~  0.00023699734~  0.000655001 \\
& & 0.0012665331~  0.0018294204~  0.0019236148~ \\
4  & 1.0 &0.0~ 0.0~ 0.0~ 0.0~ 0.0089710818~  0.028888083~  0.029891453~\\
& & -0.020672065~ -0.090109640~ -0.087959801~\\
10 & 1.0 &0.0~ 0.0~ 0.0~ 0.0~ 0.012111686~  0.027777667~ 0.018651461~\\
& & -0.018905322~ -0.049016789~ -0.033652011~\\
20 & 1.0 &0.0 ~0.0~ 0.0 ~0.0~ 0.0076518484~  0.015940557~  0.010548308~ \\
& & -0.0087795777~ -0.023213629~ -0.013514588~\\
\end{tabular}
\end{table}
}
\pagebreak
\begin{center}
	Figure Captions
\end{center}

Fig.~1 \hspace{0.6cm}
The ground state energy as a function of $U$. Filled circles are our calculated
data, with a solid line as guide to the eye.
Errorbars indicate spread of the Pad\'e approximants. Here
$E_{g}+U/4=Kinetic~Energy+U<n_{\uparrow}\;n_{\downarrow}>$.

Fig.~2 \hspace*{0.6cm}
The squared local moment $L$ as a function of $U$.
Filled circles are our calculated data, with a solid line as guide to the eye.
Errorbars indicate spread of the Pad\'e approximants.
Open squares are the results of Monte Carlo simulation by
White {\em et al.\/}~\cite{whi:89}, with dotted line as guide to the eye.
The dashed line is the result of the mean field solution.

Fig.~3 \hspace*{0.6cm}
Sublattice magnetization $2M^{+}$ as a function of $U$.
Filled circles are our calculated data, with a solid line as guide to the eye.
 Errorbars indicate spread of the Pad\'e approximants.
Long dashed line and dotted line represent mean field solution and RPA result
 respectively.  A short dashed line indicates the limiting large $U$
(Heisenberg) value.

Fig.~4 \hspace*{0.6cm}
(a) The ratio of the uniform magnetic susceptibility to $U$ as a
function of $U$.
Filled circles are our calculated data, with a dashed line as guide to the eye.
A short dashed line indicates the limiting large $U$
(Heisenberg) value.
(b) The uniform magnetic susceptibility as a function of $U$.
Filled circles are our calculated data, with a dashed line as guide to the eye.
A crossover between SDW antiferromagnetic state and Heisenberg antiferromagnet
is around $U_{c}\approx 4$.

Fig.~5 \hspace*{0.6cm}
(a) The product of the spin stiffness and $U$ as a function of $U$.
Filled circles are our calculated data, with a solid line as guide to the eye.
A short solid line indicates the limiting large $U$
(Heisenberg) value.
(b) The spin stiffness as a function of $U$.
Filled circles are our calculated data, with a solid line as guide to the eye.
Errorbars indicate spread of the Pad\'e approximants.
The solid line is the mean field result.
The filled squares are results of variational Monte Carlo
simulations~\cite{den:93}.

Fig.~6 \hspace*{0.6cm}
The spin wave velocity as a function of $U$. Filled circles are our calculated
data, with a solid line as a guide to the eye. The dotted line and dashed line
represent results of RPA and Heisenberg model respectively.

\end{document}